\documentclass[10pt,notitlepage,pra,twocolumn,showpacs]{revtex4}%
\usepackage{amsfonts}
\usepackage{amsmath}
\usepackage{amssymb}
\usepackage{graphicx}
\usepackage[toc,page,header]{appendix}%
\providecommand{\U}[1]{\protect\rule{.1in}{.1in}}

\begin{document}

\title{Anisotropic deformation of Rydberg blockade sphere in few-atom systems}

\author{Jing Qian,$^{1}$Xing-Dong Zhao,$^{2}$Lu Zhou$^{1}$ and Weiping Zhang$^{1}$}
\affiliation{$^{1}$Quantum Institute for Light and Atoms, Department of Physics, East China
Normal University, Shanghai 200062, China}
\affiliation{$^{2}$Department of Physics, Henan Normal University, Xinxiang 453007, China}

\begin{abstract}
Rydberg blockade sphere persists an intriguing picture by which a number of
collective many-body effects caused by the strong Rydberg-Rydberg interactions can
be clearly understood and profoundly investigated. In the present work, we
develop a new definition for the effective two-atom blockade radius and show that the
original spherically shaped blockade surface would be deformed when the real number
of atoms increases from two to three. This deformation of blockade sphere
reveals spatially anisotropic and shrunken properties which strongly depend
on the interatomic distance. In addition, we also study the optimal conditions
for the Rydberg antiblockade effect and make predictions for improving the
antiblockade efficiency in few-atom systems.

\end{abstract}

\maketitle

\section{Introduction}

\textit{Blockade sphere }can provide an intuitive picture to understand a number
of quantum many-body phenomena in strongly interacting Rydberg systems
\cite{Glaetzle12}, for example, the critical behavior for quantum phase
transition \cite{Weimer08}, dynamical crystallization \cite{Pohl10}, Rydberg
solitons \cite{Maucher11} and so on. Within a blockade sphere, the multi-atom excitations are significantly suppressed owing to the large energy shift
induced by the strong Rydberg-Rydberg interactions (RRIs), leading to a
collective enhancement for the single-atom transition dipole by a factor of
$\sqrt{N}$ ($N$ is the atomic number in the sphere). In a mesoscopic atomic
ensemble with $N\gg1$, the collective Rabi frequency is enhanced to $\sqrt
{N}\Omega$ and it acts as a "superatom" \cite{Heidemann07}. With this idea, a
saturable single photon absorber is designed for the manipulation of
photon-matter interactions \cite{Honer11}. The underlying physics behind
\textit{blockade sphere }is \textit{Rydberg blockade mechanism}, which was first
proposed to control quantum logic gate and generate nonclassical photonic
states \cite{Jaksch00,Lukin01}. In recent years many experimental
achievements prove the observation of single-atom collective excitation induced by the
RRIs in ultracold mesoscopic ensembles
\cite{Singer04,Dudin12,Schauss12} as well as in two atoms
\cite{Urban09,Gaetan09}. A significant reduction of the blockade size due to
the Doppler effect is observed in room temperature atomic samples
\cite{Kubler10} and can be overcome with a pulsed four-wave mixing scheme
\cite{Muller13}. These observations hold promises for further studies in
quantum information science by using strongly interacting Rydberg atoms
\cite{Saffman10,Weimer10}.

To characterize such a \textit{sphere}, defining a good blockade radius is a key
ingredient by which we can classify the whole blockade regimes into strong
blockade, partial blockade and non-blockade regimes according to the relative
strengths between the RRI energy and the linewidth of
excitation. So far, a large number of experiments and theoretical works are
performed in the strong blockade regime
\cite{Lamour08,Johnson08,Viteau11,David13} where a perfect blocked phenomenon
for the doubly excited state population is identified
\cite{Raitzsch08,Viteau12} and the formation of plasmas due to an ionization
avalanche is observed \cite{Vincent13}. More recently, L. B\'{e}guin \textit{et al.}
report a direct measurement of the van der Waals (vdWs) interaction
between two Rydberg atoms in the partial blockade regime in which the
population oscillations exhibit several vdWs energy-dependent frequencies
rather than single collective Rabi frequency \cite{Beguin13}. In order to
investigate quantum dynamics in such strong blockade and partial blockade
regimes, one always adopts a spherical-shaped model with its volume $(4/3)\pi
r_{b}^{3}$ and radius $r_{b}$ (typically a few microns) defined as the
interatomic distance at which the vdWs interaction equals the excitation
linewidth of single atom \cite{Tong04,Pritchard10,Pritchard12}. This spherical
shape assumption is robust under the condition of mean two-atom distance being
significantly smaller than $r_{b}$ in a superatom, i.e. the case of strong
blockade; however, if they two become comparable in the partial blockade
regime, we predict that the spherical surface would become deformed, leading
to an inefficient Rydberg blockade.

In a recent work, T. Pohl and P. Berman show that the two-atom Rydberg
blockade will break down when the number of atoms increases from two to three
\cite{Pohl09}. Here, we propose an alternative opinion to extend this model by 
introducing the blockade sphere picture. The
purpose of this work is to investigate the spatially anisotropic deformations
of Rydberg blockade sphere in a three-atom system where the deformation
strength is well determined by the two-atom distance. For non-interacting or
strongly interacting atoms, the two-atom blockade sphere picture is a feasible
tool to characterize and even optimize the blockade efficiency for multi-atom
ensembles. However, for partial blockade, the coherent atomic excitations
behave much complex, which makes the simple two-atom blockade sphere model 
ineffective. In addition to the complete analytical results of a two-atom system,
we numerically solve the master equations for three atoms and represent
various deformed surfaces in a two-dimensional (2D) space. Besides, we
develop an optimal antiblockade condition in few-atom systems, which is also
useful for multiple atom ensembles.

\section{A two-atom system}

\subsection{Two-atom blockade effect}

\begin{figure}[ptb]
\centering
\includegraphics[
height=3.5251in,
width=3.3264in
]{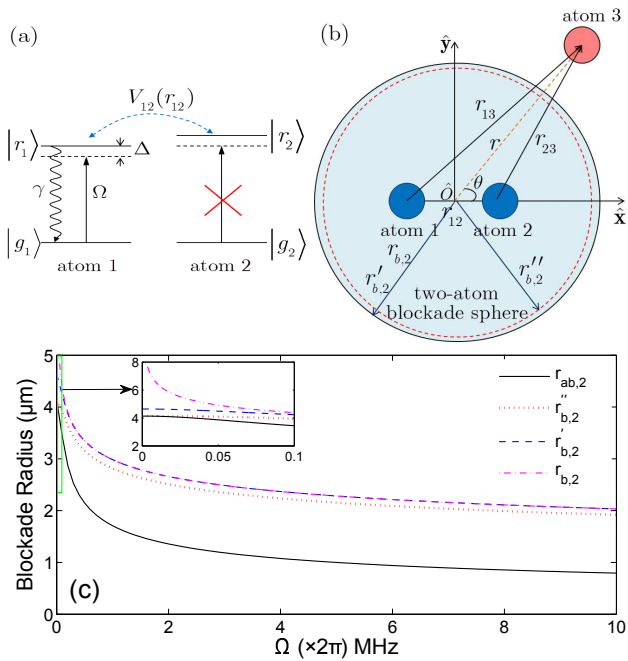}\caption{(Color online) (a) Schematic representation of two-atom 
Rydberg blockade. Each atom has a ground state $\left\vert g_{1\left(  2\right)
}\right\rangle $ and a Rydberg state $\left\vert r_{1\left(  2\right)
}\right\rangle $ which are illuminated by a cw laser field with Rabi
frequency $\Omega$ and one-photon detuning $\Delta$. $\gamma$ describes the
corresponding spontaneous decay process from the excited state to the ground
state. Due to the energy shift of state $\left\vert r_{1} r_{2}\right\rangle $ caused by
the Rydberg-Rydberg state interaction $V_{12}\left(  r_{12}\right)$, the laser is out of 
resonance for the transition from singly symmetric excited state $\left( \left\vert g_{1}r_{2}\right\rangle +\left\vert
r_{1}g_{2}\right\rangle \right) /\sqrt{2}$ to doubly excited state $\left\vert r_{1} r_{2}\right\rangle $,
and only one atom (e.g. atom 1) can be excited to $\left\vert r_{1}\right\rangle $.
(b) Two atoms 1 and 2 are
prepared with separation $r_{12}$ in a 2D space, creating a blockade sphere. A
third atom is placed in the same space, with the separations $r_{13}$ and
$r_{23}$ from atoms 1 and 2, respectively. Its position is denoted by a
controllable polar coordinate $\left(  r,\theta\right)  $ and $O$ is the
original point of the coordinate system. The radius of solid-line surface is
$r_{b,2}$ (or $r_{b,2}^{\prime}$) and of dashed-line surface is $r_{b,2}%
^{\prime\prime}$. (c) Comparing the varyings of three different two-atom
blockade radius $r_{b,2}$ (dash-dotted magenta line), $r_{b,2}^{\prime}$
(dashed blue line), $r_{b,2}^{\prime\prime}$ (dotted red line), and
one antiblockade radius $r_{ab,2}$ (solid black line) as a function of $\Omega$
values, taking into account the vdWs RRI potential energy
$V_{12}/\hbar=C_{6}/r_{12}^{6}$ for relatively low excited-state atoms with
$C_{6}\approx2\pi\times1$GHz $\mu$m$^{6}$ and the principal quantum number
$n\sim20-30$. The decay rate is
set to be $\gamma=2\pi\times0.2$MHz. Inset: a detailed picture for
$\Omega/2\pi\in\left[  0,0.1\right]  $.}%
\label{model}%
\end{figure}

Let us first recall the essence of a simple two-atom blockade scheme with
the RRI $V_{12}\left(  r_{12}\right)  $. See the
caption of Fig. \ref{model} for a detailed scheme description. When two
atoms are in the double Rydberg state $\left\vert r_{1}%
r_{2}\right\rangle $, they interact strongly due to the vdWs interaction,
leading to an interatomic distance-dependent energy shift
by $V_{12}\left(  r_{12}\right)  =C_{6}/r_{12}^{6}$ [$\hbar=1$ throughout the paper]. For the shifted energy
that is larger than the laser excitation linewidth, the laser transition 
is out of resonance and only one of two atoms (e.g. atom 1) would be
transferred to the Rydberg state at a time, resulting in very low doubly
excited state population for Rydberg atoms. This is so-called \textit{Rydberg
blockade} \textit{effect} \cite{Comparat10}.

To demonstrate the blockade efficiency quantitatively, a picture of
\textit{blockade sphere} with radius $r_{b}$ is considered to be very useful.
Within the sphere, only single-atom excitation is allowed if the
mean interatomic distance is much smaller than $r_{b}$; hence, we require a
good definition of $r_{b}$. For two atoms confined in a sphere, there are two
ways to define it:

(i) The easiest way to estimate $r_{b,2}$ (subscript $2$ means the atomic
number) relies on the model of a single laser-driven atom excited to Rydberg
state with its linewidth dominated by the power broadening $\Omega$ [$\Omega$ is single-photon Rabi frequency for two-level
atoms and effective two-photon Rabi frequency for three-level atoms], so that
$r_{b,2}$ is defined as the interatomic distance at which the interaction
energy equals the linewidth of excitation,%

\begin{equation}
r_{b,2}=\left(  \frac{C_{6}}{\sqrt{2}\Omega}\right)  ^{1/6}. \label{rb2_1}%
\end{equation}

For a $N$-atom ensemble, a collective blockade radius is $r_{b,N}=(C_{6}%
/\sqrt{N}\Omega)^{1/6}$ \cite{Heidemann07}.

(ii) Another generalized way uses the steady-state population $\left\langle
\hat{\sigma}_{rr}\right\rangle $ of a single atomic Rydberg state under cw
laser drivings that is $\left\langle \hat{\sigma}_{rr}\right\rangle
=\Omega^{2}/\left(  2\Omega^{2}+\gamma^{2}/4+\Delta^{2}\right)  $ to estimate
its linewidth $w_{1}=\sqrt{2\Omega^{2}+\gamma^{2}/4}$ (clearly $w_{1}%
\propto\sqrt{2}\Omega$) \cite{Honing13}. Therefore, the corresponding radius
$r_{b,2}^{\prime}$ takes the form of%

\begin{equation}
r_{b,2}^{\prime}=\left(  \frac{2C_{6}}{\sqrt{8\Omega^{2}+\gamma^{2}}}\right)
^{1/6}. \label{rb2_2}%
\end{equation}

So far, to our knowledge estimating blockade radius by the properties of
single atom excitation has been the unique and robust way \cite{Tong04}.
However, a full theoretical description of this laser-driven, strongly
interacting multi-atom system is still challenging. On the purpose of
improving the understandings for two-atom Rydberg blockade, we suggest a new
definition for radius by directly solving the master equations of two
interacting atoms \cite{Jing13}: $\partial_{t}\hat{\rho}=-i\left[
\mathcal{H},\hat{\rho}\right]  +\mathcal{L}\left[  \hat{\rho}\right]  $, with
the Hamiltonian $\mathcal{H}$ describing the atom-field interaction as well as
the vdWs interaction, the Lindblad operator $\mathcal{L}\left[
\hat{\rho}\right]  $ for the spontaneous decay process. Then the steady-state
population $\left\langle \hat{\sigma}_{r_{1}r_{2}}\right\rangle $ for double
Rydberg state $\left\vert r_{1}r_{2}\right\rangle $ is given by%

\begin{align}
&  \left\langle \hat{\sigma}_{r_{1}r_{2}}\right\rangle =\label{double_popu}\\
&  \frac{4\Omega^{4}}{\left(  4\Omega^{2}+\gamma^{2}/2+2\Delta^{2}\right)
^{2}+V_{12}\left(  V_{12}-4\Delta\right)  \left(  4\Omega^{2}+\gamma
^{2}/4+\Delta^{2}\right)  }.\nonumber
\end{align}

In the case of $\Delta=0$, Eq. (\ref{double_popu}) reduces into $4\Omega
^{4}/\{\left(  4\Omega^{2}+\gamma^{2}/2\right)  ^{2}+V_{12}^{2}\left(
4\Omega^{2}+\gamma^{2}/4\right)  \}$, showing a Lorentzian function of
$V_{12}$ with the two-atom excitation linewidth $w_{2}=\left(  4\Omega
^{2}+\gamma^{2}/2\right)  /\sqrt{4\Omega^{2}+\gamma^{2}/4}$ (clearly
$w_{2}\propto2\Omega$). In a similar way, we are able to define a new blockade
radius, as%
\begin{equation}
r_{b,2}^{\prime\prime}\approx\sqrt[6]{\frac{C_{6}}{w_{2}}}=\left(  \frac
{C_{6}\sqrt{16\Omega^{2}+\gamma^{2}}}{8\Omega^{2}+\gamma^{2}}\right)  ^{1/6}.
\label{new_rad}%
\end{equation}

From the definitions of radius $r_{b,2}$, $r_{b,2}^{\prime}$ and
$r_{b,2}^{\prime\prime}$, we note $r_{b,2}^{\prime\prime}<r_{b,2}^{\prime}\leq
r_{b,2}$. In Fig. \ref{model}(c), we observe that they all show a slowly
decreasing as $\Omega$ increases and $r_{b,2}^{\prime\prime}$ persists
slightly smaller than $r_{b,2}$ and $r_{b,2}^{\prime}$. $r_{b,2}$ and
$r_{b,2}^{\prime}$ have a perfect overlap if $\Omega\gg\gamma$ except for very
small $\Omega$ values [see the inset of Fig. \ref{model}(c)]. By then, the
blockade efficiency can be roughly quantified by the radius, i.e. a strong
blockade regime is defined by the two-atom distance $r_{12}$ being
significantly smaller than the blockade radius. For a large $\Omega$, the
blockade effect is weaker.

According to the results of most experiments e.g. Ref. \cite{Heidemann07},
they need the mean interaction distance being smaller than the blockade radius
by several orders of magnitude i.e. $V_{12}\gg\Omega$ for the study of strong
blockade effect, so that such differences between $r_{b,2}$, $r_{b,2}^{\prime
}$ and $r_{b,2}^{\prime\prime}$ may be negligible; however, when working in
the partial blockade regime where $V_{12}\approx\Omega$, they may become
important for the results, owing to its sensitive dependence of double
excitation probability on the interatomic distance as well as Rabi frequency
\cite{Beguin13}. In the present work, we use $r_{b,2}^{\prime\prime}$ to
represent the two-atom blockade radius.

\subsection{Two-atom antiblockade effect}

Equation (\ref{double_popu}) also provides us a new route to investigate the
antiblockade efficiency. As first predicted by C. Ates \textit{et al.}
\cite{Ate07} and later experimentally verified by T. Amthor \textit{et al.}
\cite{Amthor10}, the antiblockade effect is a specific effect with high doubly
excited state population in the blockade regime when the laser detuning
matches the Rydberg-Rydberg interaction energy. It can emerge for instance if
the RRI energy shift equals the two-photon detuning based on the
Autler-Townes splitting mechanism in an atomic three-level system. According
to the concept of antiblockade, the two-atom antiblockade condition can be
written as $2\Delta=V_{12}\left(  r\right)  $, leading to the steady-state
population $\left\langle \hat{\sigma}_{r_{1}r_{2}}\right\rangle $%

\begin{equation}
\left\langle \hat{\sigma}_{r_{1}r_{2}}\right\rangle =\frac{4\Omega^{4}%
}{\left(  4\Omega^{2}+\gamma^{2}/2\right)  ^{2}+V_{12}^{2}\gamma^{2}/4}
\label{anti_popu},%
\end{equation}
where the antiblockade radius $r_{ab,2}$ is
\begin{equation}
r_{ab,2}=\left(  \frac{\gamma C_{6}}{8\Omega^{2}+\gamma^{2}}\right)  ^{1/6}.
\label{rab2}%
\end{equation}

As shown in Fig. \ref{model}(c), compared with the blockade radius, the
antiblockade radius $r_{ab,2}$ (denoted by solid black line) is small than
$r_{b,2}^{\prime\prime}$ and has the same tendency as the blockade radius for
$\Omega$ changes.

Finally, we conclude that inside the blockade sphere, once the antiblockade
condition is met, the probability for double Rydberg excitation will be
strikingly enhanced; beyond the sphere regime, the role of RRI
is poor which makes the blockade mechanism invalid, so both high
single-atom and multi-atom excitation probabilities are possible. To observe
obvious two-atom antiblockade behavior, we suggest the interatomic distance
$r_{ab,2}<r_{12}<r_{b,2}^{\prime\prime}$ as well as the condition
$2\Delta-V_{12}\left(  r\right)  =0$ is met.

\section{A three-atom system}

\subsection{Deformed blockade sphere}

The \textit{blockade sphere} picture is identified to be robust when
demonstrating the two-atom blockade experiments \cite{Urban09,Gaetan09}, which
has extended to a $N$-atom ensemble with one excited atom, so-called
"superatom" \cite{Vuletic06}. On the other hand, while increasing the real number
of atoms from two to three, the three-atom excitation dynamics would show an
unexpected and qualitative change. This idea was considered in an isosceles
triangle configuration \cite{Pohl09} near F\"{o}rster resonance interactions
\cite{Ryabtsev10}.

In the present section, our goal is to develop a generalized model of a
three-atom system in a 2D space, for the purpose of studying spatially
anisotropic deformation of blockade sphere due to the presence of atom 3. The
assumption of a spherical-type blockade surface is widely accepted, which gives
rise to a clear picture for Rydberg studies. We show that this assumption is
exactly correct only if the separation $r_{12}$ is much smaller than the
blockade radius, otherwise, the blockade surface is no longer a spherical shape.

Our model is depicted in Fig. \ref{model}(b). Atoms 1 and 2 are initially
prepared on $x$ axis with a fixed distance $r_{12}$ and its central point $O$
describing the original point of coordinate system. In the same space, there
exists atom 3 whose position is defined as $\left(  r,\theta\right)  $ with
$r$ the distance from $O$ point and $\theta$ the directional angle with
respect to $+x$ axis. Then the interatomic separations $r_{13}$ (atom 1 and
atom 3) and $r_{23}$ (atom 2 and atom 3) can be expressed as:%

\begin{equation}
r_{13\left(  23\right)  }=\sqrt{\left(  x_{3}\pm r_{12}/2\right)  ^{2}%
+y_{3}^{2}} \label{separation}%
\end{equation}
where $\left(  x_{3},y_{3}\right)  =\left(  r/\sqrt{1+\tan^{2}\theta}%
,r\tan\theta/\sqrt{1+\tan^{2}\theta}\right)  $. Differing from a two-atom
system where only interaction $V_{12}$ plays roles, such a three-atom scheme
offers a feasible platform to study both interactions and quantum interference
from three different energy channels $V_{12}$, $V_{13}$ and $V_{23}$. In Ref.
\cite{Pohl09}, authors consider one type of configuration: an isosceles
triangle. When the third atom is brought closer to the atomic pair, by tuning the laser pulse
duration and laser amplitude, a significant increase for the double Rydberg
excitation is observed. The reason for this increase we think essentially
comes from the shrunken of blockade surface due to the multi-channel quantum interference.

In our numerical explorations, for relative low Rydberg levels (e.g.
$n\sim20-30$), the vdW coefficient is $C_{6}=2\pi\times1000$MHz $\mu$m$^{6}$
\cite{Singer05}, which for $\Omega/2\pi=1.0$MHz and $\gamma/2\pi=0.2$MHz gives
the blockade radius $r_{b,2}^{\prime\prime}=2.8155\mu$m. We note that this
value is even smaller than the collective blockade radius for three atoms
based on the former definition [$r_{b,3}=\left(  C_{6}/\sqrt{3}\Omega\right)
^{1/6}\approx2.8856\mu$m]. Besides, to perform the calculations more
efficiently, we make two assumptions: (i) If atom 3 is placed too close to
atoms 1 or 2, the interaction $V_{13}\left(  V_{23}\right)  \rightarrow
+\infty$; in this case, we treat it as a two-atom problem where the critical
distance for $r_{13}\left(  r_{23}\right)  $ is chosen as $r_{c}=1\mu$m
($1\mu$m is the characteristic wavelength of light needed for internal-state
manipulation \cite{Urban09}). (ii) By directly solving the three-atom master
equations $\partial_{t}\hat{\rho}=-i\left[  \mathcal{H},\hat{\rho}\right]
+\mathcal{L}\left[  \hat{\rho}\right]  $, we define a new quantity
$P_{2}\left(  r,\theta\right)  =\left\vert grr\right\rangle \left\langle
grr\right\vert +\left\vert rgr\right\rangle \left\langle rgr\right\vert
+\left\vert rrg\right\rangle \left\langle rrg\right\vert $ to describe the
probability of simultaneously exciting any two of three atoms to Rydberg
states. In a polar coordinate system, the three-atom blockade radius $r_{b,3}$
is numerically determined by $P_{2}\left(  r_{b,3},\theta\right)  =0.5\left(
P_{2}\left(  +\infty,\theta\right)  -P_{2}\left(  0,\theta\right)  \right)  $.
When separation $r\rightarrow+\infty$, $P_{2}$ attains its maximum for
non-blockade regimes and when $r\rightarrow0$, $P_{2}$ is minimal for strong
blockade regimes.

\begin{figure}[ptb]
\centering
\includegraphics[
height=2.9264in,
width=3.5064in
]{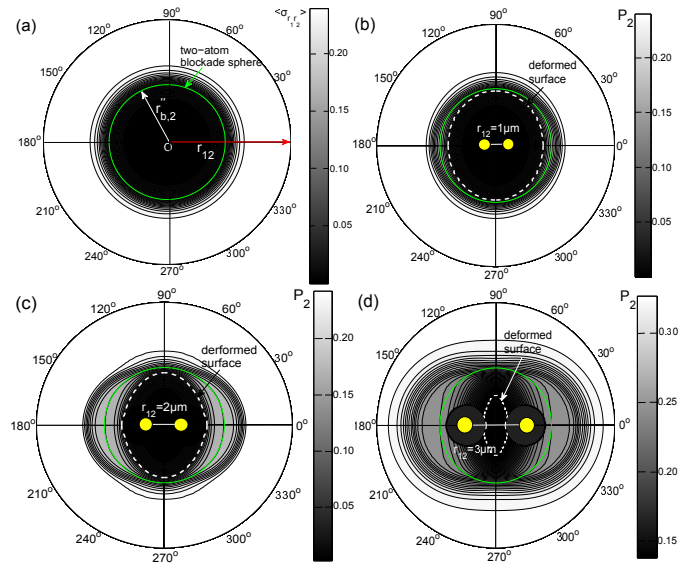}\caption{(Color online) The two-atom blockade sphere is
denoted by a solid green circle with radius $r_{b,2}^{\prime\prime}$. (a)
Along the direction of red arrow, the two-atom distance $r_{12}$ increases as
the steady-state population $\left\langle \sigma_{r_{1}r_{2}}\right\rangle $
attaining $0.25$ (white) from $0$ (black). (b)-(d) Deformed three-atom
blockade sphere surface (dashed white curve) for $r_{12}=1\mu$m, 2$\mu$m and
3$\mu$m, respectively. Two yellow circles denote atoms 1 and 2. The double
excitation probability $P_{2}$ is shown by the gray-level diagram from 0
(black) to 0.25 (white) in (b) and (c), and from 0.15 (black) to 0.35 (white)
in (d). $\Delta=0$ and other parameters are defined in the text.}%
\label{deformation}%
\end{figure}

In Fig. \ref{deformation}(a), the red arrow points to the direction of an
increasing two-atom distance $r_{12}$ along which the steady-state population
$\left\langle \hat{\sigma}_{r_{1}r_{2}}\right\rangle $ undergoes a continuous
transition from low excitations ($\left\langle \hat{\sigma}_{r_{1}r_{2}%
}\right\rangle \approx0$, black) to the saturated excitations of two-level
atoms ($\left\langle \hat{\sigma}_{r_{1}r_{2}}\right\rangle \approx0.25$,
white). The green circle denotes the two-atom blockade surface. Fig.
\ref{deformation}(b)-(d) are graphed in the 2D polar coordinate $\left(
r,\theta\right)  $ of atom 3. When atom 3 presents, the original two-atom
blockade sphere (solid green circle) reveals different strengths of
deformations, which strongly depend on the two-atom distance $r_{12}$ [see
dashed white circles]. From (b) to (d) by respectively choosing $r_{12}=1\mu
$m, 2$\mu$m and 3$\mu$m, we find the three-atom blockade spheres turn to be
radius-reducing long elliptical shape. The clear shrunken property of the
sphere surface makes the blockade effect more difficult in a three-atom
ensemble than in a two-atom one. We also find the two-atom distance $r_{12}$
has a significant role in the three-atom blockade. Compared to $r_{b,2}%
^{\prime\prime}$, the smaller $r_{12}$ is, the less broken or deformed of
blockade surface is observed while adding the third atom. Because in this
case, interactions $V_{13}$ and $V_{23}$ are always smaller or comparable with
$V_{12}$ [see Fig. \ref{deformation}(b)]. By contrast, when $r_{12}$ is close
to $r_{b,2}^{\prime\prime}$, three different interaction channels result in
strong quantum interference effect, yielding a strong deformed surface [see
Fig. \ref{deformation}(d)].

Based on our numerical results, it is interesting to find the deformation
effect in a three-atom system shows spatially anisotropic properties that the
deformed surface becomes a long ellipse rather than a circle or an oblate
ellipse. To demonstrate this finding, we consider two extreme cases: (i) a 1D
chain of atoms with $\theta=0$, (ii) an isosceles triangle with $\theta=\pi
/2$. In case (i), when atom 3 is closing to the atomic pair from $+x$ axis,
$r_{13}^{\left(  i\right)  }=r+r_{12}/2$, $r_{23}^{\left(  i\right)
}=\left\vert r-r_{12}/2\right\vert $; in case (ii), when atom 3 is coming from
$+y$ axis, $r_{13}^{\left(  ii\right)  }=r_{23}^{\left(  ii\right)  }%
=\sqrt{r_{12}^{2}/4+r^{2}}$. For given $r_{12}$ and $r$, we note that
$r_{23}^{\left(  i\right)  }<r_{13\left(  23\right)  }^{\left(  ii\right)
}<r_{13}^{\left(  i\right)  }$. A complete asymmetric excitation property for
a chain model creates a larger deformation strength than for a symmetric
isosceles triangular model; so that if $r_{12}$ is small $r_{13}^{\left(
i\right)  }\approx r_{23}^{\left(  i\right)  }$, a long elliptical-shape
surface is not so clear compared with a large $r_{12}$ case.

Another interesting result is for two-atom distance $r_{12}\left(
=3\mu\text{m}\right)  $ that is comparable with $r_{b,2}^{\prime\prime}$, as
shown in Fig. \ref{deformation}(d), the double Rydberg excitation probability
$P_{2}$ can not be absolutely suppressed even if $r\rightarrow0$. At the same
time, at $r\rightarrow+\infty$, $P_{2}$ approaches $0.35$ which is larger than
the saturation limit ($\left\langle \sigma_{r_{1}r_{2}}\right\rangle _{\max
}=0.25$) for two-level atomic systems. We think this is a regime for partial
blockade behavior (i.e. imperfect blockade) which usually appears at the
boundary of the region of finite size \cite{David13}.

\subsection{Antiblockade for multi-atom systems: take three atoms as an
example}

In section II.B, we study the two-atom antiblockade and obtain a quantified
antiblockade radius $r_{ab,2}$ for optimizing its efficiency. That is the
interatomic distance $r_{12}$ should be $r_{ab,2}<r_{12}<r_{b,2}^{\prime
\prime}$, beyond this regime, if $r_{12}<r_{ab,2}$, the antiblockade is
inefficient; otherwise if $r_{12}>r_{b,2}^{\prime\prime}$, both single and
double Rydberg states will be populated. In this section, we take a three-atom
system as an example to explore a generalized antiblockade condition form for
multi-atom systems.

In a three-atom system, we predict the antiblockade condition to be
$2\Delta-V\left(  r\right)  =0$ \cite{Li13} with $V\left(  r\right)  $ the
RRI between any two atoms $V_{12}\left(
r_{12}\right)  $, $V_{13}\left(  r_{13}\right)  $ and $V_{23}\left(
r_{23}\right)  $. First, we assume $r_{12}$ is fixed as well as $V_{12}\left(
r_{12}\right)  \neq2\Delta$, so that by changing the position of atom 3, it is
possible to meet two resonant conditions: $V_{13}\left(  r_{13}\right)
=2\Delta$ and $V_{23}\left(  r_{23}\right)  =2\Delta$ for a given detuning
$\Delta$.

\begin{figure}[ptb]
\centering
\includegraphics[
height=4.95in,
width=3.4in
]{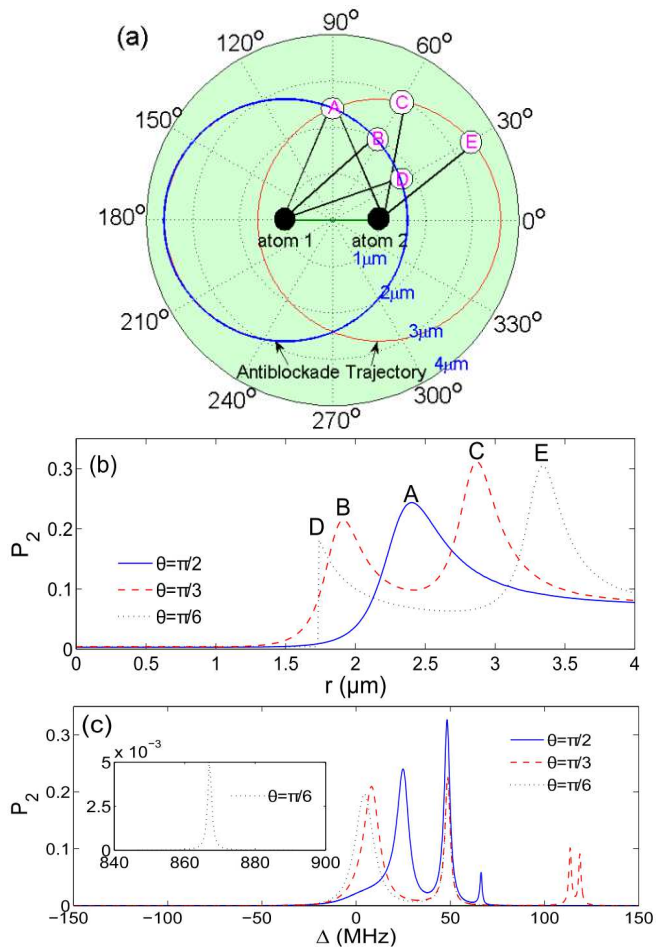}\caption{(Color online) (a) Schematic of Rydberg
antiblockade trajectory. Two crossed circles correspond to two antiblockade
conditions: $V_{13}\left(  r_{13}\right)  =2\Delta$ (left blue circle) and
$V_{23}\left(  r_{23}\right)  =2\Delta$ (right red circle). The corresponding
double Rydberg-state excitation probabilities $P_{2}$ as varied distances at
the points of A-E are plotted in (b). Parameters are $\Delta=10$MHz,
$r_{12}=2\mu$m, and others are the same as in Fig. \ref{deformation}. (c)
Probabilities $P_{2}$ vs the detuning $\Delta$ for different directional
angles $\theta=\pi/2$ (solid blue), $\pi/3$ (dashed red) and $\pi/6$ (dotted
black). }%
\label{antiblockade_pic}%
\end{figure}

In Fig. \ref{antiblockade_pic}(a), for a given detuning ($\Delta=10$MHz) and
two-atom distance ($r_{12}=2\mu$m), we show the antiblockade trajectory as two
circles (blue and red) whose centers locate at atom 1 and atom 2 and radius is
$r_{13}=r_{23}\approx2.6\mu$m, respectively to meet the resonant conditions of
$V_{13}\left(  r_{13}\right)  =2\Delta$ and $V_{23}\left(  r_{23}\right)
=2\Delta$. We select five points on two trajectories and plot their excitation
probabilities $P_{2}$ as a function of distance $r$ in Fig.
\ref{antiblockade_pic}(b), corresponding to three different angles $\theta
=\pi/2$ (A, solid blue), $\pi/3$ (B and C, dashed red), $\pi/6$ (D and E,
dotted black). The $x$-axis-labeling $r$ denotes the distance from original
point $O$ (not from atoms). From Fig. \ref{deformation}(c), we note that
points A, B, D are within the regime of deformed blockade sphere that render
the excitation probability $P_{2}\leq0.25$. However, for points C and E,
$P_{2}$ exceeds $0.25$ due to the imperfect antiblockade near the boundary of
blockade surface. In addition, we also find that except for $\theta=\pi/2$ and
$3\pi/2$ that is a symmetric triangular structure (e.g. point A), for other
angles $\theta$ there exist two distinguishable antiblockade resonances,
locating at
\begin{equation}
r=\sqrt{\left(  \frac{C_{6}}{2\Delta}\right)  ^{1/3}-\frac{r_{12}^{2}\tan
^{2}\theta}{4\left(  1+\tan^{2}\theta\right)  }}\pm\frac{r_{12}}{2\sqrt
{1+\tan^{2}\theta}} \label{resonances}%
\end{equation}

Second, we introduce another picture to study the antiblockade excitation
properties. Figure \ref{antiblockade_pic}(c) shows the double excitation
probability $P_{2}$ by varying detuning $\Delta$. We set separations
$r_{12}=2\mu$m, $r=2\mu$m, and $\theta=\pi/2$, $\pi/3$ and $\pi/6$ for three
triangular configurations. Except for $\theta=\pi/2$ that $r_{13}=r_{23}$
giving rise to two resonant excitations (denoted by solid blue lines), at
$\theta=\pi/3$ and $\pi/6$, we observe three different resonant peaks for
$P_{2}$ at $\Delta_{1}=C_{6}/2r_{12}^{6}$ (fixed, $\Delta_{1}\approx49.1$MHz),
$\Delta_{2}=C_{6}/2r_{13}^{6}$ and $\Delta_{3}=C_{6}/2r_{23}^{6}$ whose
amplitudes $P_{2}\left(  \Delta\right)  $ rely on the distance $r$. A small
$r$ value for large positive detunings inside the regime of antiblockade
radius results in a poor antiblockade efficiency [see $P_{2}\left(  \Delta
_{3}\right)  $ at $\theta=\pi/3$ and $\pi/6$]. This finding agrees well with
the two-atom case. While, at the side of negative detuning or zero detuning
values, $P_{2}\left(  \Delta\right)  $ is mostly suppressed, which strongly
verifies that the generalized antiblockade condition $2\Delta=V\left(
r\right)  $ is correct.

\section{Conclusions}

In summary, we thoroughly study the Rydberg blockade and antiblockade effects
in two- and three-atom systems. Starting from a typical two-atom blockade
sphere model, we quantify the two-atom blockade and antiblockade efficiency by
introducing more rigorous definitions for the radius, based on theoretically
solving the two-interacting-atom master equations. When turning to the
three-atom case, we observe that the spherical surface would occur strong
spatial anisotropic deformation with a reduced radius which is sensitive
related to the mean interatomic distance. This fact renders the blockade
effect more difficult to achieve in a three-atom system than in a two-atom
one. The anisotropy property stems from the asymmetric Rydberg-Rydberg
interactions for atoms 1, 3 and atoms 2, 3. In addition, we also numerically
investigate a generalized antiblockade condition in a three-atom system and
show the double Rydberg excitation properties are importantly affected by the
blockade and antiblockade boundaries.

On the one hand, our results are consistent with a collective blockade radius
$r_{b,N}\approx(C_{6}/\sqrt{N}\Omega)^{1/6}$ for a superatom scheme which
implies that $r_{b,N}$ will reduce if we increase the atomic number $N$; on
the other hand, the idea that the deformation of blockade sphere is spatially
anisotropic is brightly new and may stimulate new explorations for Rydberg
experiments in various 2D optical lattices in the future.

This work is supported by the National Basic Research Program of China (973 Program) under Grant No. 2011CB921604,
the National Natural Science Foundation of China under Grants Nos. 11104076, 11004057, 11234003 and 11374003, 
the Specialized Research Fund for the Doctoral Program of Higher Education No. 20110076120004,
the Young Scholar of Henan normal university No. 0102640065,
the 'Chen Guang' project supported by Shanghai Municipal Education Commission 
and Shanghai Education Development Foundation under grant No. 10CG24,
the Shanghai Rising-Star Program under grant No. 12QA1401000,
and the Fundamental Research Funds for the Central Universities.


\begin{thebibliography}{99}                                                                                               %


\bibitem {Glaetzle12}A. Glaetzle, R. Nath, B. Zhao, G. Pupillo and P. Zoller,
Phys. Rev. A \textbf{86} 043403(2012).

\bibitem {Weimer08}H. Weimer, R. L\"{o}w, T. Pfau and H. B\"{u}chler, Phys.
Rev. Letts. \textbf{101} 250601 (2008).

\bibitem {Pohl10}T. Pohl, E. Demler and M. Lukin, Phys. Rev. Letts.
\textbf{104} 043002 (2010).

\bibitem {Maucher11}F. Mauher, N. Henkel, M. Saffman, W. Kr\'{o}likowski, S.
Skupin and T. Pohl, Phys. Rev. Letts. \textbf{106} 170401 (2011).

\bibitem {Heidemann07}R. Heidemann, U. Raitzsch, V. Bendkowsky, B. Butscher,
R. L\"{o}w, L. Santos and T. Pfau, Phys. Rev. Letts. \textbf{99} 163601 (2007).

\bibitem {Honer11}J. Honer, R. L\"{o}w, H. Weimer, T. Pfau and H. B\"{u}chler,
Phys. Rev. Letts. \textbf{107} 093601 (2011).

\bibitem {Jaksch00}D. Jaksch, J. I. Cirac, P. Zoller, S. L. Rolston, R.
C\^{o}t\'{e} and M. D. Lukin, Phys. Rev. Letts. \textbf{85} 2208 (2000).

\bibitem {Lukin01}M. D. Lukin, M. Fleischhauer, R. C\^{o}t\'{e}, L. M. Duan,
D. Jaksch, J. I. Cirac, and P. Zoller, Phys. Rev. Letts. \textbf{87} 037901 (2001).

\bibitem {Singer04}K. Singer, M. R-Lamour, T. Amthor, L. Marcassa and M.
Weidem\"{u}ller, Phys. Rev. Letts. \textbf{93} 163001 (2004).

\bibitem {Dudin12}Y. O. Dudin and A. Kuzmich, Science \textbf{336} 887 (2012);
P. Grangier, Science \textbf{336} 812\ (2012).

\bibitem {Schauss12}P. Schauss, M. Cheneau, M. Endres, T. Fukuhara, S. Hild,
A. Omran, T. Pohl, C. Gross, S. Kuhr and I. Bloch, Nature \textbf{491} 87 (2012).

\bibitem {Urban09}E. Urban, T. A. Johnson, T. Henage, L. Isenhower, D. D.
Yavuz, T. G. Walker and M. Saffman, Nat. Phys. \textbf{5} 110 (2009).

\bibitem {Gaetan09}A. Gaetan, Y. Miroshnychenko, T. Wilk, A. Chotia, M.
Viteau, D. Comparat, P. Pillet, A. Browaeys and P. Grangier, Nat. Phys.
\textbf{5} 115 (2009).

\bibitem {Kubler10}H. K\"{u}bler, J. Shaffer, T. Baluktsian, R. L\"{o}w and T.
Pfau, Nature Phot. \textbf{4} 112 (2010).

\bibitem {Muller13}M. M\"{u}ller, A. K\"{o}lle, R. L\"{o}w, T. Pfau, T.
Calarco and S. Montangero, Phys. Rev. A \textbf{87} 053412 (2013).

\bibitem {Saffman10}M. Saffman, T. Walker and K. Molmer, Rev. Mod. Phys.
\textbf{82} 2313 (2010) and references therein.

\bibitem {Weimer10}H. Weimer, M. M\"{u}ller, I. Lesanovsky, P. Zoller and H.
B\"{u}chler, Nat. Phys. \textbf{6} 382 (2010).

\bibitem {Lamour08}M. R-Lamour, T. Amthor, J. Deiglmayr and M.
Weidem\"{u}ller, Phys. Rev. Letts. \textbf{100} 253001 (2008).

\bibitem {Johnson08}T. Johnson, E. Urban, T. Henage, L. Isenhower, D. Yavuz,
T. Walker and M. Saffman, Phys. Rev. Letts. \textbf{100} 113003 (2008).

\bibitem {Viteau11}M. Viteau, M. Bason, J. Radogostowicz, N. Malossi, D.
Ciampini, O. Morsch and E. Arimondo, Phys. Rev. Letts. \textbf{107} 060402 (2011).

\bibitem {David13}D. Petrosyan, M. H\"{o}ning and M. Fleischhauer, Phys. Rev.
A \textbf{87} 053414 (2013).

\bibitem {Raitzsch08}U. Raitzsch, V. Bendkowsky, R. Heidemann, B. Butscher, R.
L\"{o}w and T. Pfau, Phys. Rev. Letts. \textbf{100} 013002 (2008).

\bibitem {Viteau12}M. Viteau. P. Huillery, M. Bason, N. Malossi, D. Ciampini,
O. Morsch, E. Arimondo, D. Comparat and P. Pillet, Phys. Rev. Letts.
\textbf{109} 053002 (2012).

\bibitem {Vincent13}M. R-Vincent, C. Hofmann, H. Schempp, G. G\"{u}nter, S.
Whitlock and M. Weidem\"{u}ller, Phys. Rev. Letts. \textbf{110} 045004 (2013).

\bibitem {Beguin13}L. B\'{e}guin, A. Vernier, R. Chicireanu, T. Lahaye and A.
Browaeys, Phys. Rev. Letts. \textbf{110} 263201 (2013); M. Weidem\"{u}ller,
Physics \textbf{6} 71 (2013).

\bibitem {Tong04}D. Tong, S. Farooqi, J. Stanojevic, S. Krishnan, Y. Zhang, R.
Cote, E. Eyler and P. Gould, Phys. Rev. Letts. \textbf{93} 063001 (2004).

\bibitem {Pritchard10}J. Pritchard, D. Maxwell, A. Gauguet, K. Weatherill, M.
Jones and C. Adams, Phys. Rev. Letts. \textbf{105} 193601 (2010).

\bibitem {Pritchard12}J. Pritchard, C. Adams and K. Molmer, Phys. Rev. Letts.
\textbf{108} 043601 (2012).

\bibitem {Pohl09}T. Pohl and P. Berman, Phys. Rev. Letts. \textbf{102} 013004 (2009).

\bibitem {Comparat10}D. Comparat and P. Pillet, J. Opt. Soc. Am. B \textbf{27}
A208 (2010), and reference therein.

\bibitem {Honing13}M. H\"{o}ning, D. Muth, D. Petrosyan and M. Fleischhauer,
Phys. Rev. A \textbf{87} 023401 (2013).

\bibitem {Jing13}J. Qian, L Zhou and W. Zhang, Phys. Rev. A \textbf{87} 063421 (2013).

\bibitem {Ate07}C. Ates, T. Pohl, T. Pattard and J. M. Rost, Phys. Rev. Letts.
\textbf{98} 023002 (2007).

\bibitem {Amthor10}T. Amthor, C. Giese, C. Hofmann and M. Weidem\"{u}ller,
Phys. Rev. Letts. \textbf{104} 013001 (2010).

\bibitem {Vuletic06}V. Vuletic, Nature Phys. \textbf{2} 801 (2006).

\bibitem {Ryabtsev10}I. Ryabtsev, D. Tretyakov, I. Beterov, V. Entin and E.
Yakshina, Phys. Rev. A \textbf{82} 053409 (2010).

\bibitem {Singer05}K. Singer, J. Stanojevic, M. Weidem\"{u}ller and R.
C\^{o}t\'{e}, J. Phys. B: At. Mol. Opt. Phys. \textbf{38} S295 (2005).

\bibitem {Li13}W. Li, C. Ates and I. Lesanovsky, Phys. Rev. Letts.
\textbf{110} 213005 (2013).
\end{thebibliography}
\end{document}